\documentclass[10pt,twocolumn,letterpaper]{article}

\usepackage{iccv}
\usepackage{times}
\usepackage{epsfig}
\usepackage{graphicx}
\usepackage{amsmath}
\usepackage{amssymb}
\usepackage{footnote}
\usepackage{url}
\makesavenoteenv{tabular}
\makesavenoteenv{table}

\usepackage{subeqnarray}
\usepackage{multirow}
\usepackage{tabularx}
\usepackage{epstopdf}
\usepackage{subfigure}
\usepackage{array}

\newcolumntype{I}{!{\vrule width 3pt}}
\newlength\savedwidth

\newlength\savewidth

\newcolumntype{C}{>{\centering\arraybackslash}X}
\usepackage[pagebackref=true,breaklinks=true,letterpaper=true,colorlinks,bookmarks=false]{hyperref}

\usepackage{cleveref}

\iccvfinalcopy 

\def\iccvPaperID{****} 

\ificcvfinal\fi
\begin{document}
\vspace{-0.2cm}
\title{AIM 2019 Challenge on Constrained Super-Resolution: Methods and Results}

\author{Kai Zhang
\and Shuhang Gu
\and Radu Timofte
\and Zheng Hui
\and Xiumei Wang
\and Xinbo Gao
\and Dongliang Xiong
\and Shuai Liu
\and Ruipeng Gang
\and Nan Nan
\and Chenghua Li
\and Xueyi Zou
\and Ning Kang
\and Zhan Wang
\and Hang Xu
\and Chaofeng Wang
\and Zheng Li
\and Linlin Wang
\and Jun Shi
\and Wenyu Sun
\and Zhiqiang Lang
\and Jiangtao Nie
\and Wei Wei
\and Lei Zhang
\and Yazhe Niu
\and Peijin Zhuo
\and Xiangzhen Kong
\and Long Sun
\and Wenhao Wang
}

\maketitle
\thispagestyle{empty}

\begin{abstract}
This paper reviews the AIM 2019 challenge on constrained example-based single image super-resolution with focus on proposed solutions and results.
The challenge had 3 tracks. Taking the three main aspects (\ie, number of parameters, inference/running time, fidelity (PSNR)) of MSRResNet as the baseline, Track 1 aims to reduce the amount of parameters while being constrained to maintain or improve the running time and the PSNR result, Tracks 2 and 3 aim to optimize running time and PSNR result with constrain of the other two aspects, respectively. Each track had an average of 64 registered participants, and 12 teams submitted the final results. They gauge the state-of-the-art in single image super-resolution.
\end{abstract}
\section{Introduction}
\let\thefootnote\relax\footnotetext{K. Zhang (kai.zhang@vision.ee.ethz.ch, ETH Zurich), S. Gu, and R. Timofte (timofter@vision.ee.ethz.ch, ETH Zurich) are the challenge organizers, while the other authors participated in the challenge.
\\Appendix~\ref{sec:teams} contains the authors' teams and affiliations.
\\AIM webpage: \url{http://www.vision.ee.ethz.ch/aim19/}}

Image super-resolution (SR) is a classical but still active and challenging research topic in computer vision due to its ill-poseness nature and high practical values~\cite{Yang-CVPR-2008,Timofte-ACCV-2014,dong2014learning,Kim-CVPR-2016,wang2018esrgan,zhang2019deep,li2019_3dappearance}.
The goal of SR is to recover a visually pleasant high-resolution (HR) version of a low-resolution (LR) input.

Existing researches on SR mainly focus on example-based methods, among which convolutional neural networks (CNN) have shown unprecedented success in effectiveness and efficiency. In order to benchmark CNN-based SR, a series of SR challenges have been successfully organized.
In NTIRE 2017~\cite{Timofte_2017_CVPR_Workshops,Agustsson_2017_CVPR_Workshops}, the SR challenge provided a large DIV2K~\cite{Agustsson_2017_CVPR_Workshops} dataset and focused on SR with standard bicubic degradation and `unknown' (blur and decimation) degradations. Among the challenge solutions, EDSR~\cite{lim2017enhanced} achieves excellent results, demonstrating that deeper and wider network can significantly improve the PSNR performance. In NTIRE 2018~\cite{Timofte_2018_CVPR_Workshops}, the SR challenge considered $\times8$ SR with bicubic degradation and $\times$4 SR with unknown synthetic degradations. The challenge promoted realistic degradation settings and introduced several novel SR architectures~\cite{DBPN2018,zhang2018learning} with promising results. In PIRM 2018, the SR challenges focused on improving the perceptual quality of super-resolved images~\cite{Blau_2018_ECCV_Workshops} and efficiency on smartphones~\cite{Ignatov_2018_ECCV_Workshops}. ESRGAN~\cite{wang2018esrgan}, as one of the outstanding solutions, has attracted increasing attentions due to its superior performance over the GAN-based SR work SRGAN~\cite{ledig2017photo}. In NTIRE 2019~\cite{cai2019ntire}, the SR challenge aimed to explore solutions for real image SR based on real LR/HR training pairs. Undoubtedly, most of the challenge solutions rely on very deep architectures.
Although the above challenges continuously advanced the improvement of SR from various aspects, the challenge solutions generally achieve top results with the sacrifice of efficiency, and it is difficult to make a fair comparison for those solutions. From the perspective of real applications, a CNN-based SR solution should be fast and effective with limited computational resources.

Jointly with AIM 2019 workshop we have an AIM challenge on constrained super-resolution, that is, the task of super-resolving an input image to an output image with a magnification factor $\times$4 while the proposed solution is constrained on any two of the following aspects: number of parameters, running (inference) time, fidelity (PSNR). The challenge attempts to provide a relatively fair comparison for the challenge solutions, and it has three tracks, each of which focuses on one of the aspects.
This challenge is expected to be a step forward in benchmarking example-based single image super-resolution.
In the next, we will describe the challenge, present and discuss the results and describe the methods.

\section{AIM 2019 Challenge}
The objectives of the AIM 2019 challenge on constrained super-resolution challenge are:
(i) to explore advanced SR with potentially balanced trade-off between number of parameters, effectiveness and efficiency; (ii) to compare different solutions under a properly fair setting; and (iii) to offer an opportunity for academic and industrial attendees to interact and explore collaborations.

\vspace{-0.1cm}
\subsection{DIV2K Dataset~\cite{Agustsson_2017_CVPR_Workshops}}
Following~\cite{Agustsson_2017_CVPR_Workshops}, the DIV2K dataset which contains 1,000 DIVerse 2K resolution RGB images is employed in our challenge. The HR DIV2K is divided into 800 training images, 100 validation images and 100 testing images.
The corresponding LR DIV2K in this challenge is the bicubicly downsampled counterpart with scale factor 4.
Note that the testing HR images are completely hidden from the participants during the whole challenge.
In order to get access to the data and submit the testing HR results, registration on Codalab (\url{https://competitions.codalab.org/}) is required.

\vspace{-0.05cm}
\subsection{MSRResNet Baseline Model~\cite{wang2018esrgan}}
We choose the MSRResNet with 16 residual blocks as the baseline model.
Different from the original SRResNet~\cite{ledig2017photo}, MSRResNet made following two modifications.
First, each residual block of MSRResNet consists of two $3\times3$ convolutional layers with Leaky ReLU activation in the middle and an identity skip connection summed to its output.
Second, the global identity skip connection directly sums the bilinearly interpolated LR image to the output of final convolutional layer. It is worth pointing out that MSRResNet used DIV2K~\cite{Agustsson_2017_CVPR_Workshops}, Flickr2K and OST~\cite{wang2018recovering} datasets for training. The quantitative measures of MSRResNet are given as follows: (1) the number of parameters is 1,517,571 (1.5M); (2) the average PSNRs on validation and testing sets of DIV2K are 29.00 dB and 28.70dB, respectively; (3) the average running time over validation and testing sets with PyTorch 1.1.0, CUDA Toolkit 10.0, and a single Titan Xp GPU are 0.130 seconds and 0.125 seconds, respectively.

\vspace{-0.05cm}
\subsection{Tracks and Competitions}

\vspace{0.07cm}
\noindent{\textbf{Track 1: Parameters}}, the aim is to obtain a network design with the lowest amount of parameters while being constrained to maintain or improve the PSNR result and the running time of MSRResNet.

\vspace{0.07cm}
\noindent{\textbf{Track 2: Inference}}, the aim is to obtain a network design with the lowest inference running time on a common GPU (\eg, Titan Xp) while being constrained to maintain or improve over MSRResNet in terms of the number of parameters and the PSNR result.

\vspace{0.07cm}
\noindent{\textbf{Track 3: Fidelity}}, the aim is to obtain a network design with the best fidelity (PSNR) while being constrained to maintain or improve over MSRResNet in terms of the number of parameters and the running time on a common GPU (\eg, Titan Xp).

\vspace{0.07cm}
\noindent{\textbf{Challenge phases }}
\textit{(1) Development phase:} the participants got the 800 LR/HR training image pairs and 100 LR validation images of the DIV2K dataset; the participants got the MSRResNet model from github (\url{https://github.com/znsc/MSRResNet}) and then obtained the baseline running time on their own computer.
\textit{(2) Validation phase:}
the participants uploaded the HR validation results, self-reported number of parameters and self-reported running time to an online validation server to get immediate feedback.
During this phase, the ranks of three tracks were based on the self-reported number of parameters, self-reported running time, and the PSNR results, respectively.
\textit{(3) Testing phase:} the participants got 100 HR validation images and 100 LR testing images; the participants submitted their super-resolved results to Codalab and emailed the code and factsheet to the organizers;
the organizers checked and ran the provided code to get the final number of parameters, running time and PSNR value for all 3 tracks; the participants got the final results at the end of the challenge.

\vspace{0.07cm}
\noindent{\textbf{Evaluation protocol }}
The quantitative measures are number of parameters, average running time and average PSNR. For the average running time, it is evaluated over the 100 LR validation images rather than the LR testing images for convenience. In addition, the best average running time among three consecutive trails is selected as the final result.
For the average PSNR, it is evaluated over the 100 testing LR images. As a common setting in SR literature~\cite{Timofte_2017_CVPR_Workshops}, a boundary of $4$ image pixels are ignored for both estimated HR results and its ground-truth HR counterpart in calculating PSNR.

\begin{table*}[!htbp]\footnotesize 
\caption{AIM 2019 constrained SR challenge results and final rankings for \textbf{Track 1: Parameters}.} 
\center
\begin{tabular}{p{1.5cm}|p{1.7cm}||p{0.9cm}<{\centering}|r|p{1.7cm}<{\centering}||p{1.7cm}<{\centering}|p{1.7cm}<{\centering}|p{1.6cm}<{\centering}|p{0.7cm}<{\centering}}
 \multirow{3}{*}{Team} &  \multirow{3}{*}{Author}  & \multirow{3}{*}{PSNR} & Number\quad\quad & Running & Running & Running time & \multirow{3}{*}{GPU} & \multirow{2}{*}{Extra} \\
  &  & &  of\quad\quad~ & time (seconds)  & time (seconds)  & (seconds) of& & \multirow{2}{*}{data}\\
    &  & &  parameters~~ & by organizers & by authors  & MSRResNet &  &      \\\hline\hline
rainbow	&zheng222&	28.78	&893,936$_{(1)}$	&0.055	&0.039	&0.091	&RTX 2080Ti	&Yes\\

Alpha &	q935970314&	28.71	&1,127,064$_{(2)}$&	0.084&	0.066&	0.097&	RTX 2080Ti&	Yes\\

ZJUCSR2019	&BearMaxZJU	&28.73	&1,185,219$_{(3)}$	&0.121	&0.145	&0.150	&Titan Xp&	No\\

Rookie	&WangChaofeng	&28.81&	1,387,258$_{(4)}$	&0.121	&0.065	&0.101	&RTX 2080Ti	&No\\

krahaon\_ai\_cv&	Krahaon\_ai\_cv&28.84&1,461,735$_{(5)}$	&0.117	&0.110	&0.112	&Titan Xp&	Yes\\
\hline
Baseline	&MSRResNet&	28.70&	1,517,571$_{(6)}$&0.130	&0.130	&0.130	&Titan Xp&	Yes\\
\hline
SRSTAR	&wy-sun16	&28.65	&852,874$_{(-)}$	&0.126	&0.130	&0.136	&RTX 2080	&Yes\\

NPUCS\_103&	Lab\_103&	28.66	&910,467$_{(-)}$	&0.123	&0.131	&0.132	&GTX 1080Ti	&No\\

PPZ&	PaParaZz1&	28.57	&818,432$_{(-)}$&	0.100	&0.082	&0.112	&Titan Xp	&No\\
neptuneai	&neptuneai	&28.88&	1,204,227$_{(-)}$	&0.452 &0.720	&0.520	&Tesla P100	&Yes\\
GUET-HMI	&suen	&28.54	&536,005$_{(-)}$&	0.290&	0.399	&0.579	&Tesla P100&	No\\

\end{tabular}
\label{table_track1}
\end{table*}

\begin{table*}[!htbp]\footnotesize 
\caption{AIM 2019 constrained SR challenge results and final rankings for \textbf{Track 2: Inference}.} 
\center
\begin{tabular}{p{1.5cm}|p{1.7cm}||p{0.9cm}<{\centering}|r|p{1.7cm}<{\centering}||p{1.7cm}<{\centering}|p{1.7cm}<{\centering}|p{1.6cm}<{\centering}|p{0.7cm}<{\centering}}
 \multirow{3}{*}{Team} &  \multirow{3}{*}{Author}  & \multirow{3}{*}{PSNR} & Number\quad\quad & Running & Running & Running time & \multirow{3}{*}{GPU} & \multirow{2}{*}{Extra} \\
   &  & &  of\quad\quad~ & time (seconds)  & time (seconds)  & (seconds) of& & \multirow{2}{*}{data}\\
    &  & &  parameters~~ & by organizers & by authors  & MSRResNet &  &      \\\hline\hline

rainbow	&zheng222	&28.78	&893,936&	0.055$_{(1)}$	&0.039	&0.091	&RTX 2080Ti&	Yes\\
ZJUCSR2019&	BearMaxZJU	&28.73	&1,227,340&	0.066$_{(2)}$&	0.081	&0.150	&Titan Xp&	No\\
Alpha	& q935970314	&28.71	&1,127,064&	0.084$_{(3)}$& 0.066	&0.097	&RTX 2080Ti	&Yes\\
krahaon\_ai\_cv	&Krahaon\_ai\_cv	&28.84	&1,461,735	&0.117$_{(4)}$	&0.110	&0.112&	Titan Xp	&Yes\\
Rookie&	WangChaofeng	&28.81	&1,387,258&	0.121$_{(5)}$	&0.065	&0.101&	RTX 2080Ti&	No\\
SRSTAR	&wy-sun16	&28.69	&1,074,447	&0.129$_{(6)}$	&0.122	&0.136	&RTX 2080	&Yes\\
\hline
Baseline	&MSRResNet&	28.70&	1,517,571&0.130$_{(7)}$	&0.130	&0.130	&Titan Xp&	Yes\\
\hline
GUET-HMI	&suen	&28.54	&536,005&	0.290$_{(-)}$	&0.399	&0.579	&Tesla P100	&No\\
neptuneai	&neptuneai&	28.88	&1,204,227&	0.452$_{(-)}$&	0.720	&0.520	&Telsa P100&	Yes\\

\end{tabular}
\label{table_track2}
\end{table*}

\begin{table*}[!htbp]\footnotesize 
\caption{AIM 2019 constrained SR challenge results and final rankings for \textbf{Track 3: Fidelity}.} 
\center
\begin{tabular}{p{1.5cm}|p{1.7cm}||p{0.9cm}<{\centering}|r|p{1.7cm}<{\centering}||p{1.7cm}<{\centering}|p{1.7cm}<{\centering}|p{1.6cm}<{\centering}|p{0.7cm}<{\centering}}
 \multirow{3}{*}{Team} &  \multirow{3}{*}{Author}  & \multirow{3}{*}{PSNR} & Number\quad\quad & Running & Running & Running time & \multirow{3}{*}{GPU} & \multirow{2}{*}{Extra} \\
   &  & &  of\quad\quad~ & time (seconds)  & time (seconds)  & (seconds) of& & \multirow{2}{*}{data}\\
    &  & &  parameters~~ & by organizers & by authors  & MSRResNet &  &      \\\hline\hline
krahaon\_ai\_cv	&Krahaon\_ai\_cv	&28.84$_{(1)}$	&1,461,735	&0.117	&0.110&	0.112	&Titan Xp	&Yes\\
Rookie	&WangChaofeng&	28.81$_{(2)}$	&1,387,258	&0.121	&0.065	&0.101	&RTX 2080Ti&	No\\
rainbow	&zheng222	&28.78$_{(3)}$	&893,936&	0.055	&0.039&	0.091&	RTX 2080Ti	&Yes\\
ZJUCSR2019	&BearMaxZJU&	28.78$_{(4)}$&	1,480,643	&0.125	&0.149	&0.150&	Titan Xp&	No\\
SRSTAR	&wy-sun16	&28.74$_{(5)}$	&1,369,859	&0.142&	0.137	&0.136	&RTX 2080	&Yes\\
Alpha	&q935970314	&28.71$_{(6)}$	&1,127,064&	0.084	&0.066	&0.097	&RTX 2080Ti	&Yes\\
\hline
Baseline	&MSRResNet&	28.70$_{(7)}$&	1,517,571&0.130	&0.130	&0.130	&Titan Xp&	Yes\\
\hline
neptuneai	&neptuneai	&28.88$_{(-)}$	&1,204,227	&0.452	&0.720&	0.520	&Tesla P100	&Yes\\
GUET-HMI	&suen	&28.54$_{(-)}$	&536,005&	0.290	&0.399	&0.579	&Tesla P100	&No\\
\end{tabular}
\label{table_track3}\vspace{-0.1cm}
\end{table*}

\vspace{-0.05cm}
\section{Challenge Results}
From 64 registered participants on average per each track, 12 teams entered in the final phase and submitted results, codes/executables, and factsheets. Because 2 teams, DeepSR and IPCV\_IITM, proposed to use more parameters than baseline, their results are unfortunately excluded in this report. Tables~\ref{table_track1}-\ref{table_track3} report the final test results and rankings of the challenge. The solutions with worse fidelity (PSNR), inference time, or number of parameters than the MSRResNet baseline are not ranked.
The methods are briefly described in section~\ref{sec:methods_and_teams} and the team members are listed in Appendix~\ref{sec:teams}.

\vspace{0.08cm}
\noindent{\textbf{Architectures and main ideas }}
Most of the proposed methods try to improve MSRResNet via various ways, such as modifying residual blocks, changing activation function, enforcing parameter and feature reuse, reducing channel number and using dilated convolution. Some typical methods are given in the following.
The overall first place winner \textit{rainbow} proposed a new residual block which enforces feature reuse, adopts 1$\times$1 convolution, and has a large receptive field of two conventional 3$\times$3 residual blocks. As such, the method of rainbow reduces the number of residual blocks from 16 to 8, thus achieving an excellent balance among number of parameters, inference speed, PSNR performance and memory occupation. Rather than adopting hand-designed architecture, krahaon\_ai\_cv and neptuneai proposed to use neural architecture search to obtain their best models.
The method of PPZ fuses different tricks, such as squeeze-excitation block, h-swish and h-sigmoid activation function and classic channel pruning method, to improve the baseline MSRResNet.
With a deep analysis on the building blocks of MSRResNet, ZJUCSR2019 proposed a new efficient upsampling block to reduce inference time.

\vspace{0.08cm}
\noindent{\textbf{Track 1: Parameters }}
The GUET-HMI team reported the lowest number of parameters (\ie, 536,005), but it is inferior to MSRResNet on PSNR and running time. While rainbow, SRSTAR and NPUCS\_103 have similar number of parameters, rainbow is the one which outperforms MSRResNet in PSNR on testing set and is the fastest among all the entries.

\vspace{0.08cm}
\noindent{\textbf{Track 2: Inference }}
The rainbow team achieves the lowest running time, an average of 0.055s over the 100 validation LR images on Titan Xp GPU. It is worth noting that, among the top 3 methods on this track, rainbow also outperforms the other two methods with respect to number of parameters and PSNR performance.

\vspace{0.08cm}
\noindent{\textbf{Track 3: Fidelity }}
Compared to krahaon\_ai\_cv, neptuneai has a better PSNR performance and lower number of parameters, however neptuneai reported that it has a slower inference than baseline MSRResNet and the running time tested by organizers on Titan Xp is 0.452s, much slower than baseline 0.130s. Rookie comes 0.03dB behind krahaon\_ai\_cv with fewer parameters.

\vspace{0.08cm}
\noindent{\textbf{Ensembles }}
Since the running time of each solution is constrained and the commonly-used model-ensemble and self-ensemble~\cite{Timofte-CVPR-2016}
would increase the running time, none of the top ranked teams employ such a strategy to improve the PSNR performance.

\vspace{0.08cm}
\noindent{\textbf{Train data and loss function }}
Following baseline MSRResNet which used DIV2K, Flickr2K and OST datasets for training, nearly half methods, such as the methods proposed by rainbow, Alpha, and krahaon\_ai\_cv, also adopted Flickr2K as additional training dataset. In addition, flipping and rotation based data augmentation~\cite{Timofte-CVPR-2016} was utilized for most of the methods.
For the loss function, most of the methods adopted the L1 loss function. Especially noteworthy is that NPUCS\_103 team proposed L1 based focal loss to enhance the PSNR performance.

\vspace{0.08cm}
\noindent{\textbf{Conclusions }}
From the above analysis of different solutions, we can have several conclusions.
(i) The proposed methods improve the state-of-the-art in balancing the trade-off between number of parameters, running time and PSNR results for SR.
(ii) It is possible to design a network with much fewer parameters and much less running time than MSRResNet without sacrifice of PSNR results. (iii) Existing commonly-used residual blocks with two $3\times3$ convolutional layers are redundant, an information multi-distillation block (IMDB) can potentially replace two such residual blocks to reduce the number of parameters and improve the inference speed without losing accuracy. (iv) Apart from hand-designed architecture, neural architecture search and model pruning are good alternative solutions for fast and effective SR.

\section{Challenge Methods and Teams}
\label{sec:methods_and_teams}


\subsection{rainbow Team}
\begin{figure}[!h]
    \centering
    \includegraphics[width=.75\linewidth]{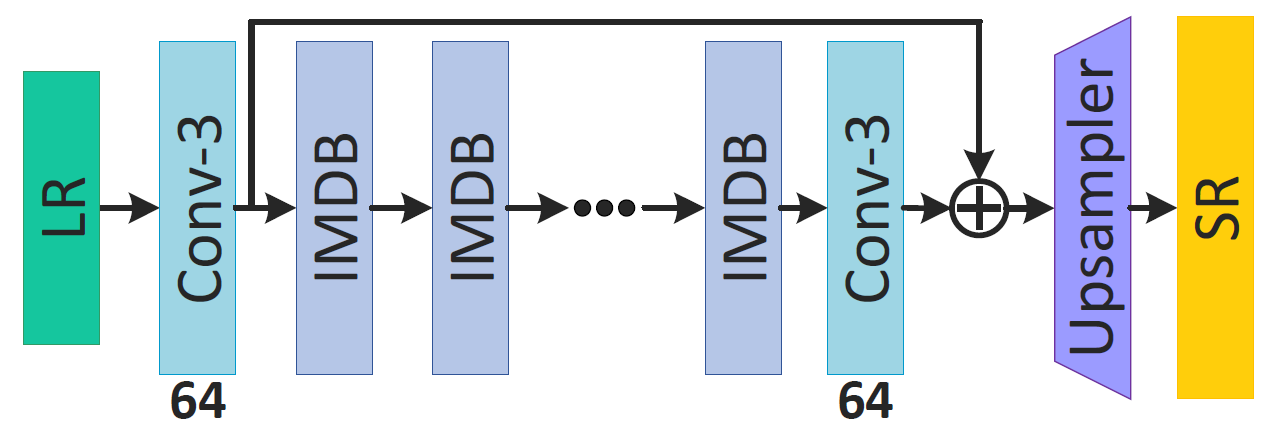}
    \caption{rainbow Team: architecture of information multi-distillation neetwork (IMDN).}
    \label{fig:rainbow-1}
    \vspace{-0.2cm}
\end{figure}

\begin{figure}[!h]
    \centering
    \includegraphics[width=.45\linewidth]{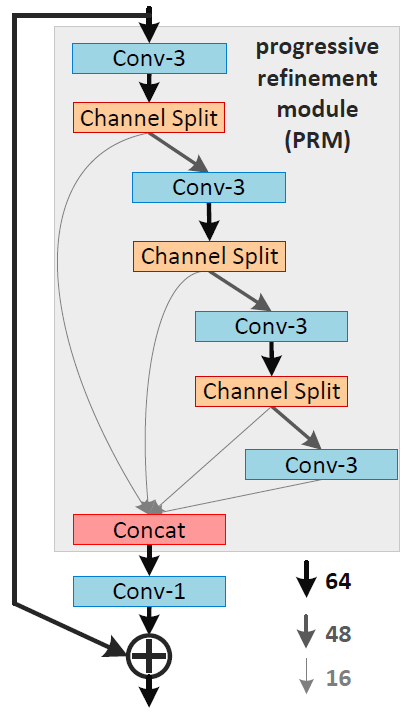}
    \caption{rainbow Team: architecture of the information multi-distillation block (IMDB). Here, 64, 48 and 16 all represent the number of output channels. ``Conv-3" denotes the $3\times3$ convlution layer.}
    \label{fig:rainbow-2}
    \vspace{-0.2cm}
\end{figure}

The rainbow team proposed \textbf{information multi-distillation network (IMDN)} for the three tracks. The proposed IMDN, as illustrated in Figure~\ref{fig:rainbow-1}, is inspired by their published works~\cite{hui2018fast,hui2019lightweight}.
The main idea of IMDN is to replace the 16 residual blocks of MSRResNet with 8 information multi-distillation modules (IMDB).  The IMDB block is illustrated in Figure~\ref{fig:rainbow-2}, from which we can see that IMDB extracts hierarchical features step-by-step, and then aggregates them by simply using a 1$\times$1 convolution. Thanks to the split operations, the input channels are reduced to achieve excellent balance among the number of parameters, inference speed, PSNR performance, and memory occupation.
During the training of IMDN, HR patches of size 640$\times$640 are randomly cropped from HR images, and the mini-batch size is set to 25. The IMDN model is trained by minimizing L1 loss function with Adam optimizer. The initial learning rate is set to $2\times10^{-4}$ and halved at every $1.42\times10^{5}$ iterations. Apart from DIV2K dataset, Flickr2K dataset is also used for the training of IMDN.

\subsection{ZJUCSR2019 Team}
The ZJUCSR2019 team proposed \textbf{efficient neural network for super-resolution without upsampling convolution (NoUCSR)}~\cite{Xiong-ICCVW-2019} for the three tracks.

\begin{figure}[!h]
    \centering
    \includegraphics[width=\linewidth]{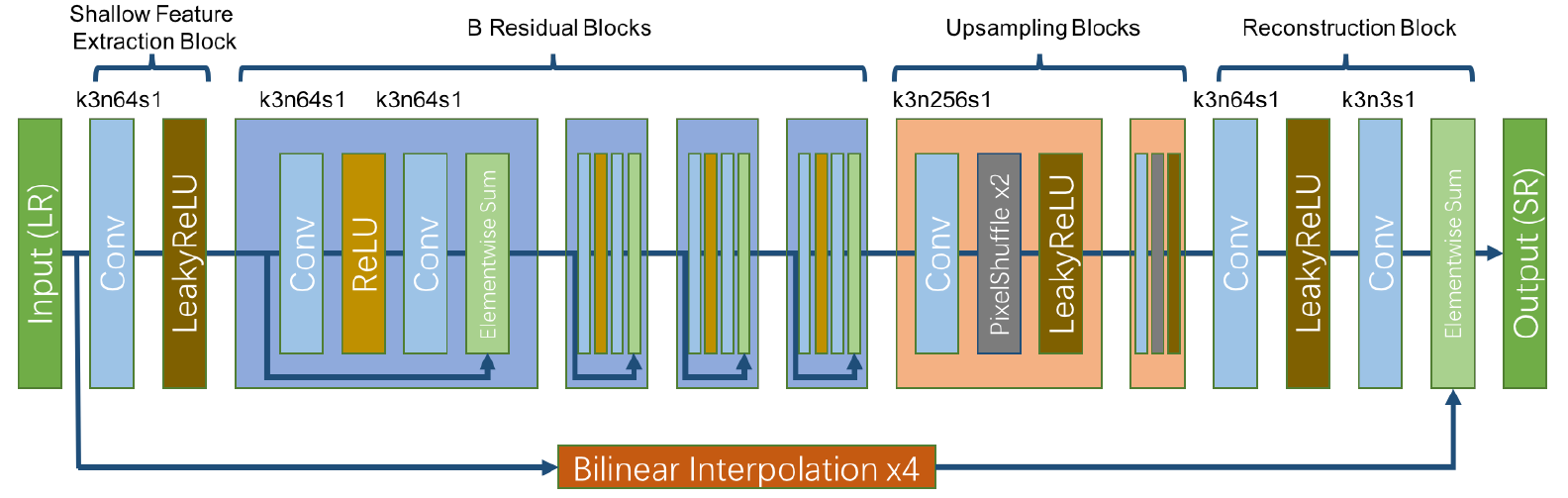}
    \vspace{-2mm}
    \caption{ZJUCSR2019 Team: MSSRResNet baseline architecture.}
    \label{fig:ZJUCSR2019-1}
    \vspace{-4mm}
\end{figure}

\begin{figure}[!h]
    \centering
    \includegraphics[width=.75\linewidth]{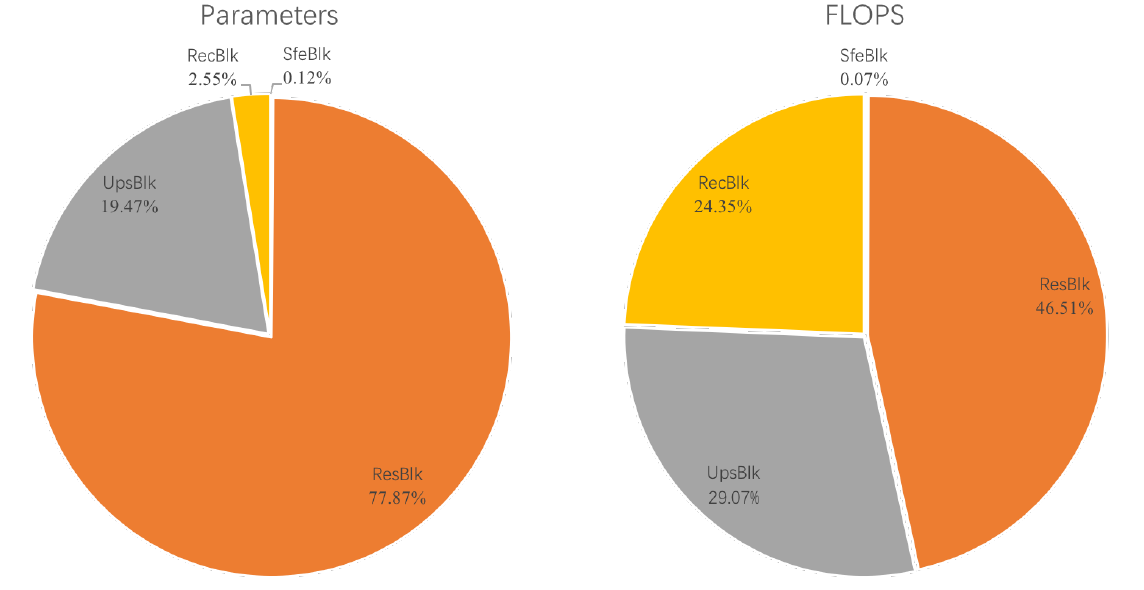}
    \vspace{-2mm}
    \caption{ZJUCSR2019 Team: parameters and FLOPs ratio of each block in MSRResNet.}
    \label{fig:ZJUCSR2019-2}
    \vspace{-4mm}
\end{figure}

First, the analyses on the architecture of baseline MSRResNet are provided.
As shown in Figure \ref{fig:ZJUCSR2019-1}, the baseline MSRResNet consists of four parts: shallow feature extraction block (SfeBlk), $B=16$ residual blocks (ResBlk), upsampling blocks (UpsBlk) and reconstruction block (RecBlk). Figure \ref{fig:ZJUCSR2019-2} shows the parameters and FLOPs ratio of each block in MSRResNet. For parameters, Resblk with 16 residual blocks contributes 77.87\%, while UpsBlk with only 2 upsampling convolution layers contributes 19.47\%. For FLOPs, SfeBlk, Resblk, UpsBlk and RecBlk contributes 0.07\%, 46.51\%, 29.07\%, and 24.35\% respectively. To reduce inference time, one can reduce the computation in RecBlk as much as possible. If high output channels are provided for pixel-shuffle~\cite{shi2016real} layer without upsampling convolution layers, then parameters and inference time of UpsBlk can be reduced significantly.

\begin{figure}[!h]
    \centering
    \includegraphics[width=\linewidth]{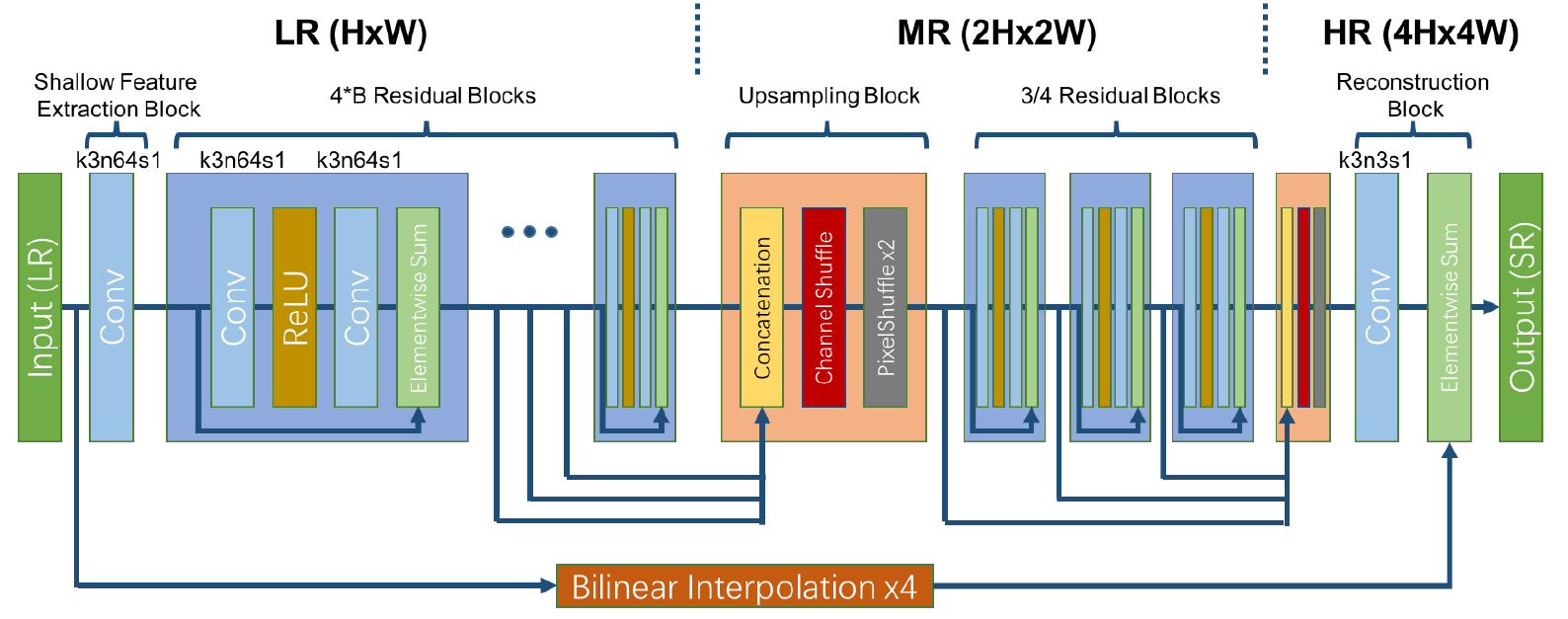}
    \vspace{-2mm}
    \caption{ZJUCSR2019 Team: architecture of NoUCSR network.}
    \label{fig:ZJUCSR2019-3}
    \vspace{-4mm}
\end{figure}

Second, the upsampling convolution layers are replaced by concatenating different level features \cite{tong2017image}, and one convolution layer at most in RecBlk is used to
reduce inference time. To maintain performance, progressive upsampling
architecture is adopted. Since the main contribution is to replace upsampling convolution, the proposed network illustrated in Figure \ref{fig:ZJUCSR2019-3} is called NoUCSR.

\subsection{Alpha Team}
The Alpha team proposed \textbf{Aggregative Structure in Super Resolution (ASSR)}~\cite{ASSR19} for the three tracks. ASSR supplements a lightweight guideline for SR networks based on the design criteria of the four lightweight networks proposed by shufflenetv2~\cite{ma2018shufflenet}.
The four criteria are given as follows: (1)~same channel width can minimize the memory access cost; (2)~excessive group convolution will increase the cost of memory access; (3)~internal fragmental operations in the network will reduce the degree of parallelism; (4)~Bias, ReLU, AddTensor, and Depthwise Convolution operations do not add too much FLOPs, but increase the MAC.
ASSR basically satisfies the design guidelines of shufflenetv2 because it uses a same channel width through the entire network, basically deletes all the bias in the network, does not split feature map and use any special operations that can reduce network parameters such as Depthwise Convolution and Pointwise Convolution.

\begin{figure}[!h]
	\centering
	\includegraphics[width=\linewidth]{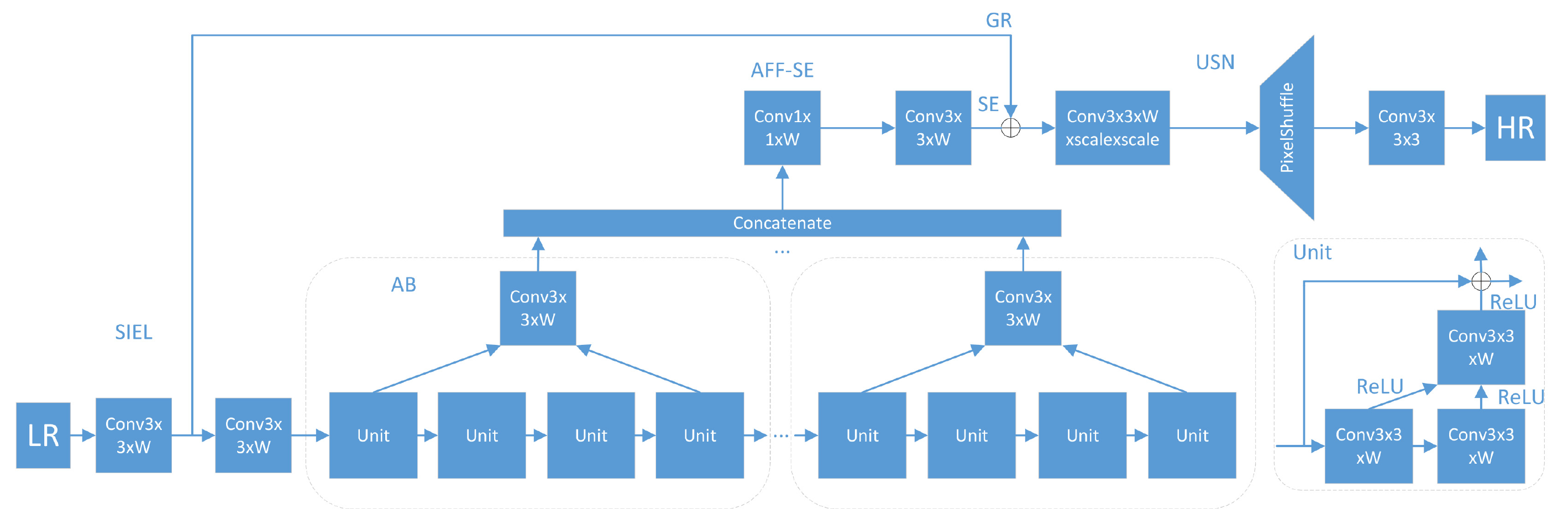}
	\caption{Alpha Team: architecture of ASSR.}\label{fig:Alpha-1}
    \vspace{-3mm}
\end{figure}

The proposed ASSR, as illustrated in Figure~\ref{fig:Alpha-1}, mainly includes the following contents: Shallow Information Extraction Layer (SIEL), Aggregative Block (AB), Aggregative Feature Fusion with Squeeze-and-Excitation (AFF-SE), Global Residual (GR) and Up-Sampling Net (USN). The SIEL is used to extract shallow representation information. Each AB has 4 Aggregative Residual units~\cite{yu2018deep} which has only one concatenate operation. In AFF-SE, a 1$\times$1 convolution operation is used to linearly combine features after concatenating. Then, an SE module that can explicitly model the interdependencies between feature channels in the convolution layer is followed. The SE module consists of two operations: Squeeze and Excitation. The squeeze operation compresses all 2-dimensional feature channels of the convolution layer into global 1-dimensional values by global average pooling to obtain an output vector with global corresponding features.
After getting the output, the Global Residual is adopted to give the network overall guidance, thus making the subject network more sparse.
For USN, it includes the expansion convolution operation and the PixelShuffle operation~\cite{shi2016real}.

\subsection{krahaon\_ai\_cv Team}
The krahaon\_ai\_cv team proposed \textbf{efficient super-resolution network with neural architecture search} for the three tracks.
The proposed method is based on neural architecture search. It uses relatively large network width in the early stage of the network and then gradually reduces the width to reduce the number of parameters and inference time.

After analyzing the baseline MSRResNet, it is argued that the upconv layers and all layers after are too computationally heavy (activation functions are ignored here):
conv3$\times$3(64, 256) $\rightarrow$ PixelShuffle(2) $\rightarrow$ conv3$\times$3(64, 256) $\rightarrow$ PixelShuffle(2) $\rightarrow$ conv3$\times$3(64, 64) $\rightarrow$ conv3$\times$3(64, 3).
In MSRResNet, the channel size remains 64 when the resolution increases to 16 times that of the original.
Obviously, if the number of channels is reduced, the network can have a PSNR performance gain by adding more residual blocks. The corresponding part of MSRResNet is modified to:
$n_x$ residual blocks $\rightarrow$ conv3$\times$3($x$, $y\times4$) $\rightarrow$ PixelShuffle(2) $\rightarrow$ $n_y$ residual blocks $\rightarrow$ conv3$\times$3($y$, $z\times4$) $\rightarrow$ PixelShuffle(2) $\rightarrow$ $n_z$ residual blocks $\rightarrow$ conv3$\times$3($z$, 3).
While the baseline MSRResNet is similar to the case where $x = y = z = 64$, $n_x = 16$, $n_y = n_z = 0$, the proposed method performs a random search on these variables. Specifically, it chooses $x$ from $\{48, 64, 80, 96\}$, $y$ from $\{x/2, x/4\}$, $z$ from $\{y/2, y/4\}$, $n_x$ from $[10, 30]$, $n_y$ from $[0, 16]$ and $n_z$ from $[0, 16]$.

After training about 300 randomly generated models with about 1 day on a machine with 8 tesla V100 GPUs, the model with $x = 64$, $y = 16$, $z = 4$, $n_x = 19$, $n_y = 12$, $n_z = 3$ achieves the highest PSNR result and is selected as the final network solution.
Then a fine-tuning stage with increased batch size of 64 from 16 is performed.
Finally, Flickr2K dataset and a small learning rate of $2\times10^{-5}$ are further utilized to have a PSNR gain of 0.16dB on DIV2K validation set.

\subsection{Rookie Team}
\begin{figure}[!h]
    \centering
    \includegraphics[width=\linewidth]{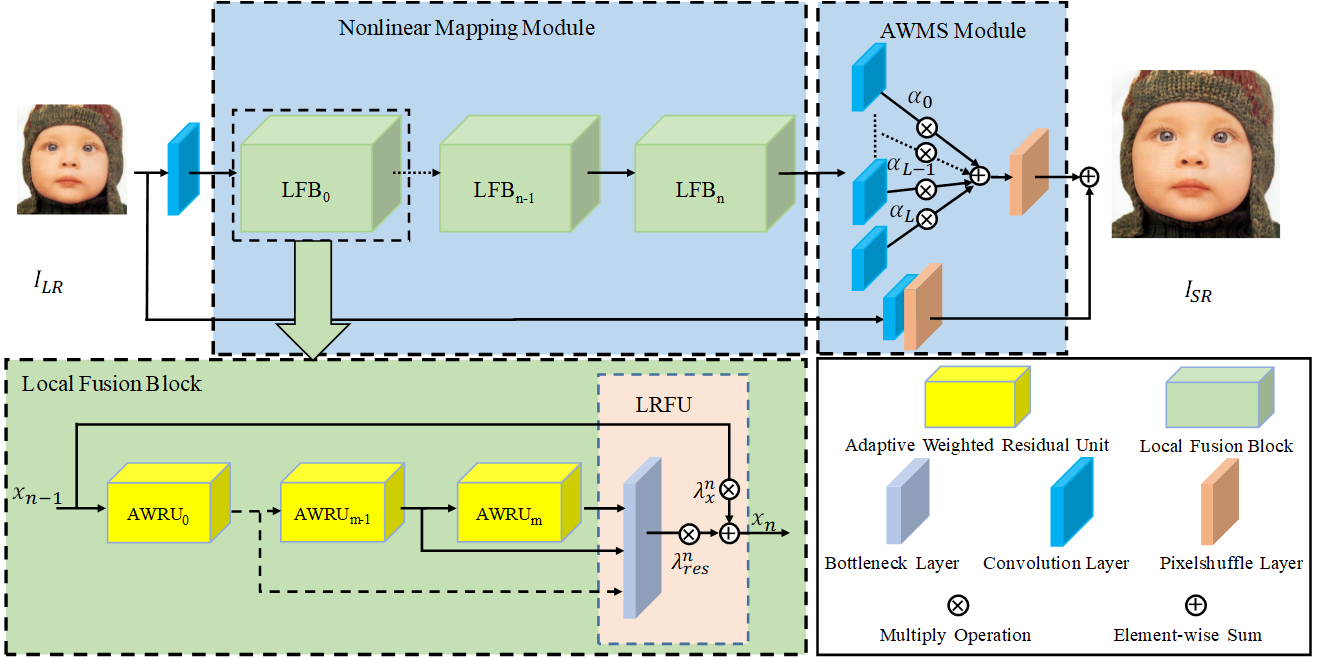}
    \vspace{-1mm}
    \caption{Rookie Team: architecture of AWSRN.}
    \label{fig:Rookie-1}
    \vspace{-4mm}
\end{figure}

\begin{figure}[!h]
    \centering
    \includegraphics[width=.65\linewidth]{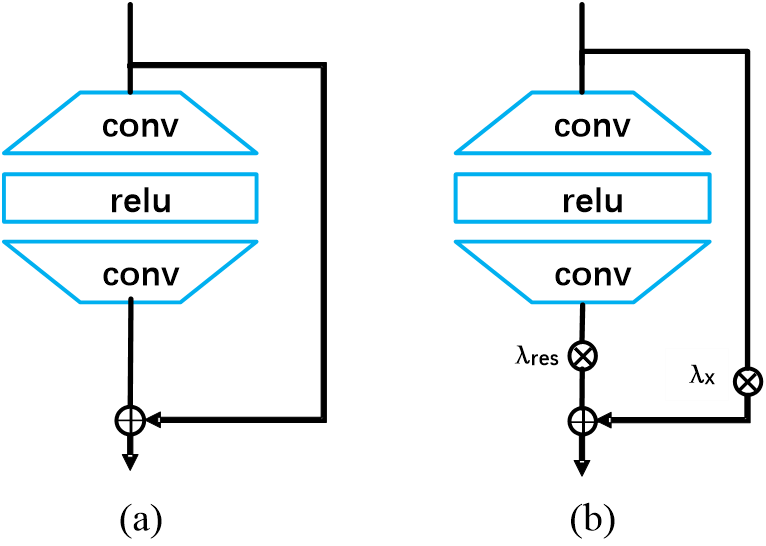}
    \vspace{-1mm}
    \caption{Rookie Team: (a) basic RU from WDSR without any weights; (b) AWRU that has two independent weights for the residual unit.}
    \label{fig:Rookie-2}
    \vspace{-4mm}
\end{figure}

The Rookie team proposed \textbf{Adaptive Weighted Super-Resolution Network (AWSRN)} for the three tracks.
As illustrated in Figure~\ref{fig:Rookie-1}, a novel local fusion block (LFB) which consists of stacked adaptive weighted residual units (AWRU) and a local residual fusion unit (LRFU) is designed for efficient residual learning. In addition, an adaptive weighted multi-scale (AWMS) module is proposed to make full use of features in reconstruction layer. AWMS consists of several different scale convolutions, and the redundancy scale branch can be removed according to the contribution of adaptive weights in AWMS for lightweight network. Moreover, the wide-activate residual unit in WDSR~\cite{yu2018wide} shown in Figure \ref{fig:Rookie-2}(a) is employed as the basic residual unit (Basic RU). Such unit allows more low-level information to be activated without increasing parameters by shrinking the dimensions of the input/output and extending the internal dimensions before ReLU. Based on the Basic RU, AWRU which additionally contains two learnable parameters (see Figure \ref{fig:Rookie-2}(b)) is then proposed. Specially,
the two learnable parameters of AWRU can be regarded as a generalized form of residual scaling~\cite{lim2017enhanced}.

\subsection{SRSTAR Team}
\begin{figure}[!h]
    \centering
    \includegraphics[width=.8\linewidth]{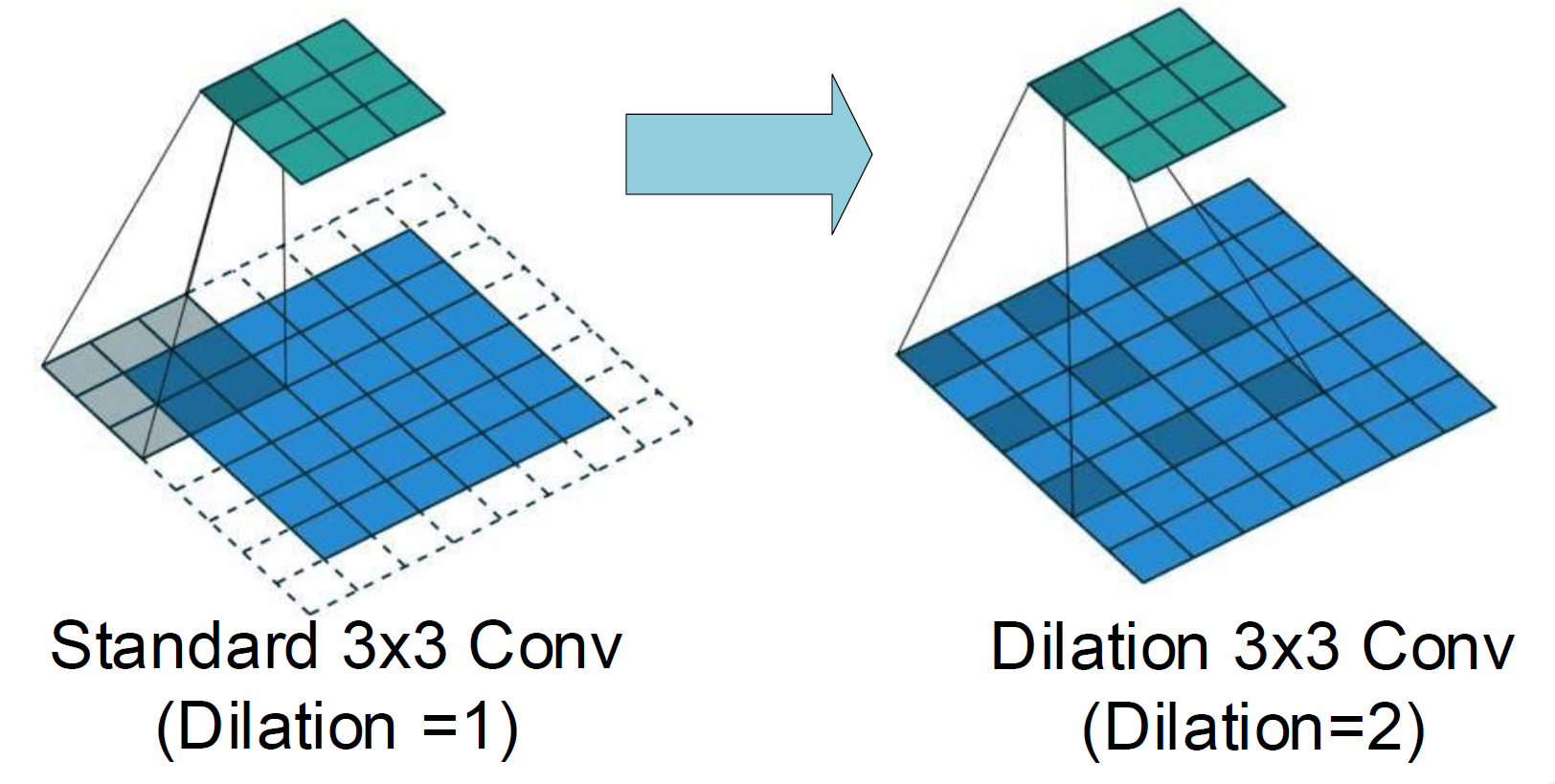}
    \vspace{-1mm}
    \caption{SRSTAR Team: dilation convolutions in DilaResNet.}
    \label{fig:SRSTAR-1}
    \vspace{-2mm}
\end{figure}

\begin{figure}[!h]
    \centering
    \includegraphics[width=\linewidth]{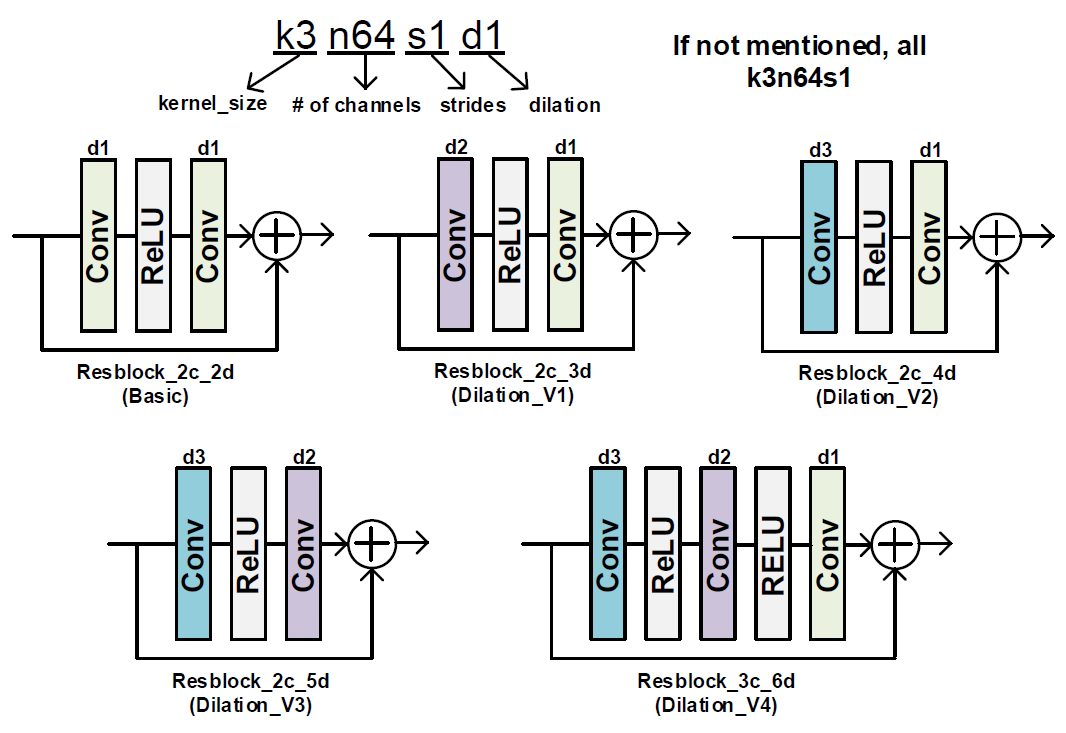}
    \vspace{-4mm}
    \caption{SRSTAR Team: residual blocks in DilaResNet.}
    \label{fig:SRSTAR-2}
    \vspace{-2mm}
\end{figure}

The SRSTAR team proposed \textbf{DilaResNet} and its variants for the three tracks.
Based on the main structure of SRResNet. DilaResNet adopts dilation convolution to increase the receptive fields of network when number of convolution layers is limited. The kernel size of all convolution layers is 3$\times$3 as it is efficient to expand receptive fields and reduce computing time in PyTorch. In addition, the network with dilation convolution can have larger receptive field, which can enable larger training patch sizes for better performance. Figure~\ref{fig:SRSTAR-1} illustrates the difference between standard convolution and dilated convolution with dilation factor 2. Moreover, residual scaling is deployed for the ultra large patch size with deep networks to make the network focus on the center receptive fields.
As shown in Figure \ref{fig:SRSTAR-2}, several residual blocks with different receptive fields (owing to different dilation factor setup) are defined to improve the performance with constrained number of convolutions. In the three tracks, different combinations of residual blocks are proposed to achieve the best performance.

For track 2, the number of residual convolution is decreased
from 32 of SRResNet baseline to 24 of DilaResNet. Two kinds of DilaResdual blocks are utilized.  The inference time is decreased from 0.13s per images of MSRResNet to 0.11s per images of DilaResNet with a single RTX 2080 GPU.  Residual scaling is not used because of shallow depth of the network.

For track 3, following the same idea of track 2, the number of residual convolution is decreased from 32 to 30 and a residual scaling of 0.5 is used. Furthermore, more kinds of DilaResdual blocks are introduced to enhance the expressive ability of network for better PSNR. The final model achieves an average PSNR of 29.10 dB over the DIV2K validation dataset.

For track 1, following the same idea of track 3, the number of residual convolution is decreased from 32 to 30 and 0.5 residual scaling is used. In order to reduce the number of parameters, some (dilation) convolutions and some DilaResDual block share same parameters.  The parameters decreases from the baseline 1,517,571 to 798,339.

During the training of DilaResNet, the HR patch size is set to 320$\times$320 in order to utilize more spatial information. The cosine annealing learning scheme rather than the multi-step scheme is adopted since it has a faster training speed.
The initial maximum learning rate is set to 2e-4 and the minimum learning rate is set to 1e-7. The period of cosine is 250k iterations and maximum learning rate decays to 1/4 at each periods.
Flickr2K dataset is also used to train the models.

\subsection{NPUCS\_103 Team}
\begin{figure}[!h]
    \centering
    \includegraphics[width=\linewidth]{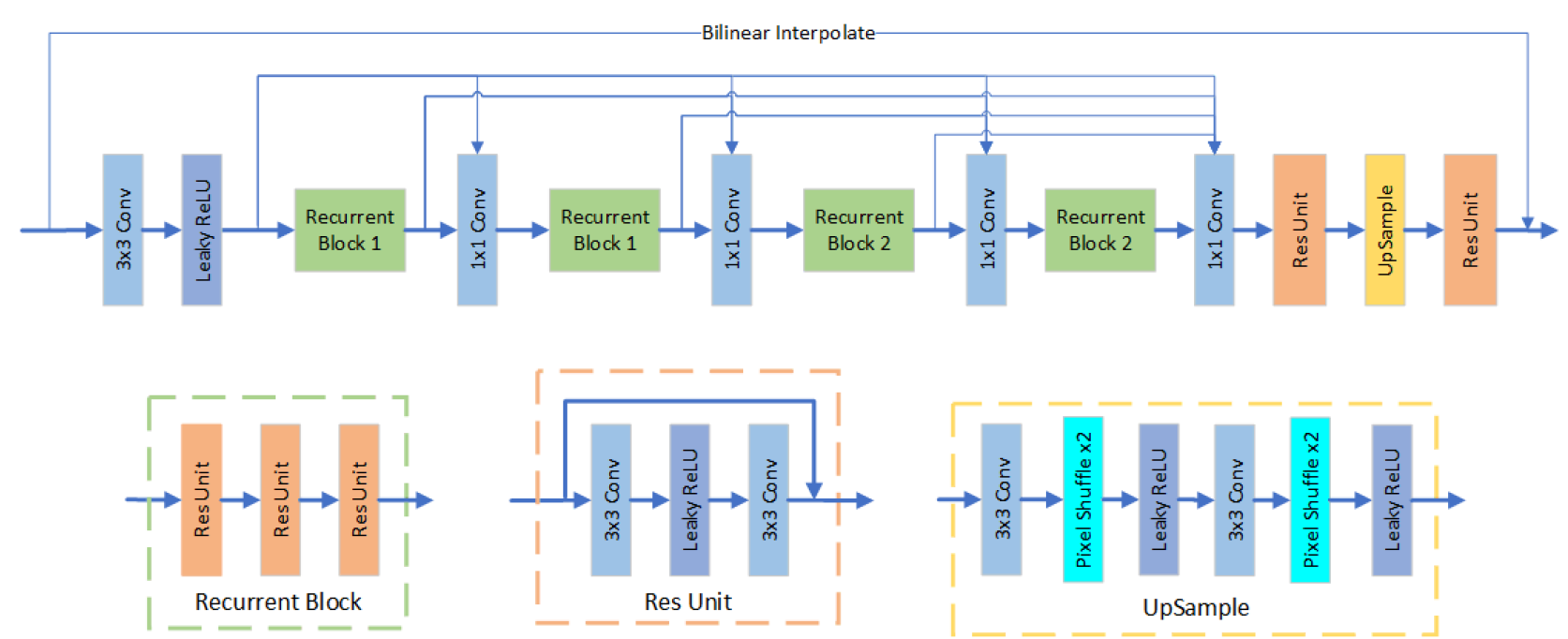}
    \vspace{-1mm}
    \caption{NPUCS\_103 Team: architecture of two-stage recurrent dense network.}
    \label{fig:NPUCS_103-1}
    \vspace{-4mm}
\end{figure}

The NPUCS\_103 team proposed a \textbf{two-stage recurrent dense net} (see Figure \ref{fig:NPUCS_103-1}) for track 1. To reduce the parameters, a recurrent architecture which contains two parts with independent circulation is adopted as the backbone net. In addition, the modules are organized into a dense connection mode to improve the PSNR performance. Moreover, a L1 based focal loss is adopted to further improve the PSNR performance.
Different from MSRResNet which employs 16 residual modules, the proposed method only utilizes 6 residual modules. The 6 residual modules are divided into two parts, \ie, RecurrentBlock\_1 and RecurrentBlock\_2. Each part is repeatedly used twice, and the output feature map of each recurrent stage with the feature map of last stage are concatenated as a dense connection.
To reduce the dimension of concatenated feature maps, an extra $1\times1$ convolution layer behind RecurrentBlock\_1 and RecurrentBlock\_2 is added.
To rectify the output, a fusion layer before up-sample layer is inserted.
As for the focal loss, the difference between super-resolved image and ground truth is used as a weight of L1 loss, thus making the model focuses on images with large reconstruction errors.

\subsection{PPZ Team}
\begin{figure}[t]
    \centering
    \includegraphics[width=\linewidth]{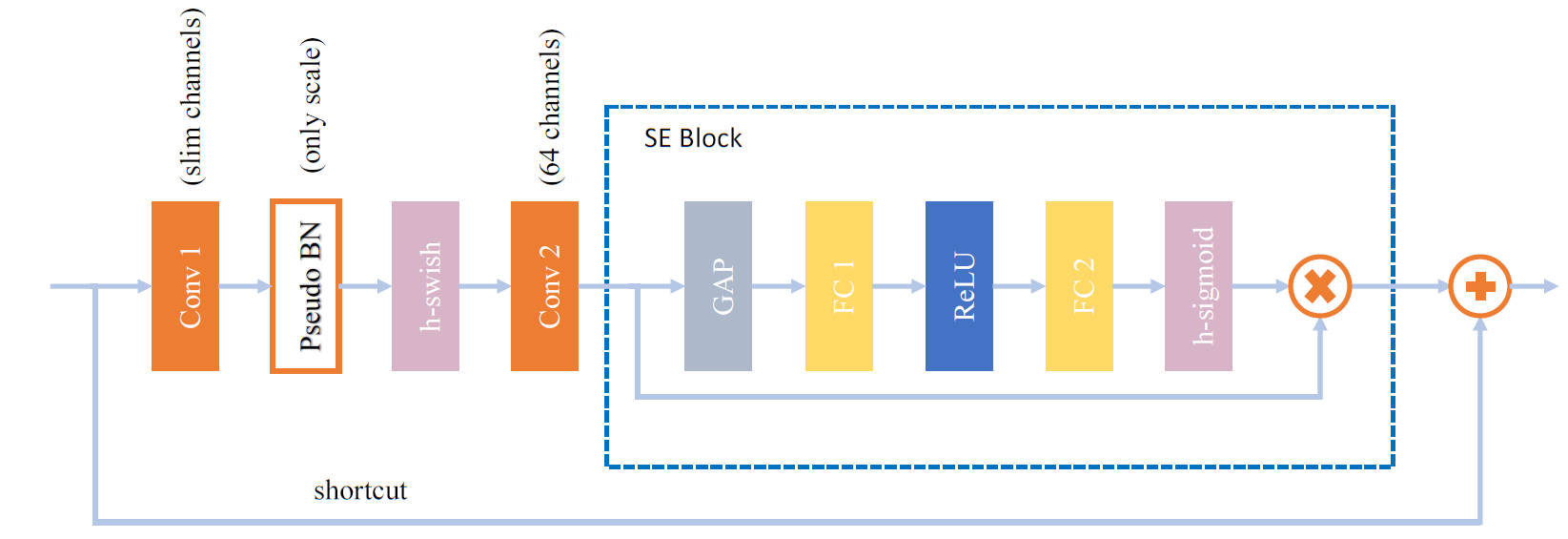}
    \vspace{-1mm}
    \caption{PPZ Team: block overview. The proposed network uses no-bias and reflect-pad convolution module as basic operation, combines SE block with the classic ResNet block and replace the activation by h-swish and h-sigmoid.}
    \label{fig:PPZ-1}
    \vspace{-4mm}
\end{figure}

The PPZ team proposed \textbf{Bags of Tricks for the Parameter-Efficient SRResNet} for track 1.
The proposed method is based on the following observations and analyses.
First, it is empirically shown that even if there are only 10 blocks, the PSNR of MSRResNet drops a little (about 0.05dB). Second, in terms of network structure, a FALSR like model is considered, but the final benefit is limited. Third, more complicated bi-cubic interpolation rather than the simple bi-linear interpolation is expected to improve the PSNR performance. Fourth,  each block's sub-structure can be optimized by following modifications (see Figure~\ref{fig:PPZ-1}): (1) adding Squeeze-Excitation structure to enhance the feature extraction ability of block; (2) adopting h-swish and h-sigmoid \cite{howard2019searching} activation function to improve the PSNR performance and relieve overhead; (3) using reflect padding to replace default zero padding in order to improve the result of image edge part; (4) removing all bias of convolution in feature extraction module; (5) employing classic channel pruning method \cite{liu2017learning} to reduce channel number of the first convolutional layer of each block.

For model pruning, the proposed method adopts the network-slimming \cite{liu2017learning} method to reduce the parameters of the model. Network-slimming is based on Batch Normalization's Learning-Able weight. As a consideration of channel redundancy, before pruning begins, the network is trained to calculate L1 norm as a penalty for the weight, so as to improve the sparsity of the model (the corresponding BN weight will be close to 0), pruning operations will remove a certain percentage of low-weight channels. Then a fine-tuning stage is performed for the pruned network to complete the network-slimming.
It turns out that even after pruning on the next 10 blocks, one can still prune nearly 35\% of the channel without losing accuracy. The model is trained with Adam by minimizing L1 loss and feature mimic loss. The learning rate, patch size, and batch size are set to 1e-4, 196 and 16, respectively.

\subsection{neptuneai Team}
\begin{figure}[t!]
    \centering
    \includegraphics[width=.75\linewidth]{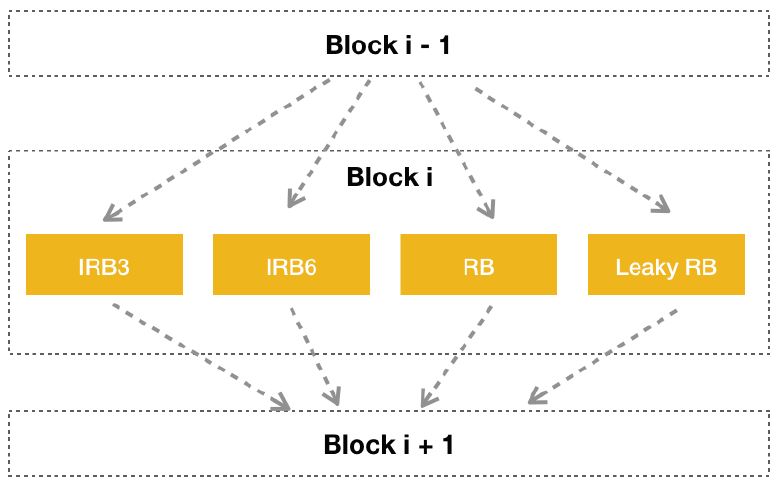}
    \vspace{-1mm}
    \caption{neptuneai Team: basic blocks for NAS.}
    \label{fig:neptuneai-2}
    \vspace{-4mm}
\end{figure}

\begin{figure}[t!]
    \centering
    \includegraphics[width=\linewidth]{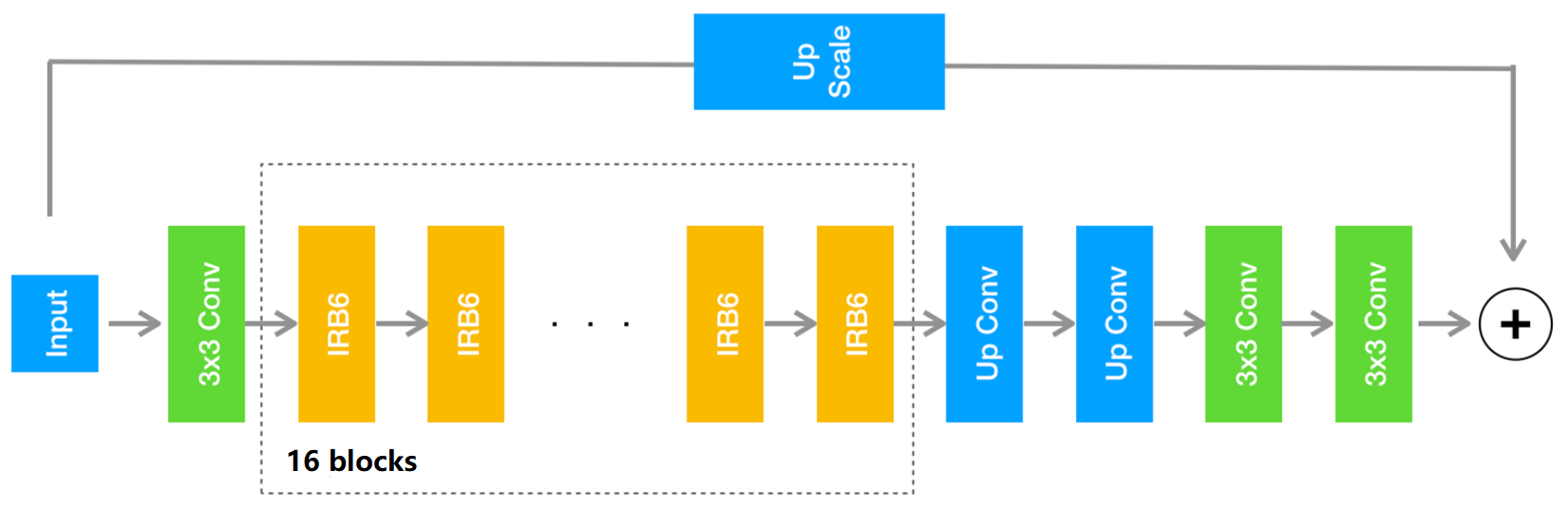}
    \vspace{-3mm}
    \caption{neptuneai Team: architecture of baseline model.}
    \label{fig:neptuneai-1}
    \vspace{-4mm}
\end{figure}

The neptuneai team proposed \textbf{lightweight super resolution network with inverted residuals blocks}
for the three tracks. The main idea is to change the basic blocks of MSRResNet to Inverted Residuals Block from MobileNetv2. By first designing a search space which is composed of several kind of blocks, the Neural Architecture Search (NAS) algorithm is adopted to discover the best architecture for light weight super resolution tasks.
The basic blocks for NAS (see Figure \ref{fig:neptuneai-2}) includes: 1)~Inverted Residuals Block with 3 expand ration, 2)~Inverted Residuals Block with 6 expand ration, 3)~Basic Residual Block and 4)~Basic Residual Block with leaky ReLU.
With NAS, a good lightweight network architecture is found. However, the re-training of this network is not done yet. Thus, the current performance of the best architecture is below than the baseline model. All the submissions are generated with the baseline model with inverted residuals blocks. Figure~\ref{fig:neptuneai-1} illustrates the architecture of baseline model.

\subsection{GUET-HMI Team}
\begin{figure}[t!]
    \centering
    \includegraphics[width=.9\linewidth]{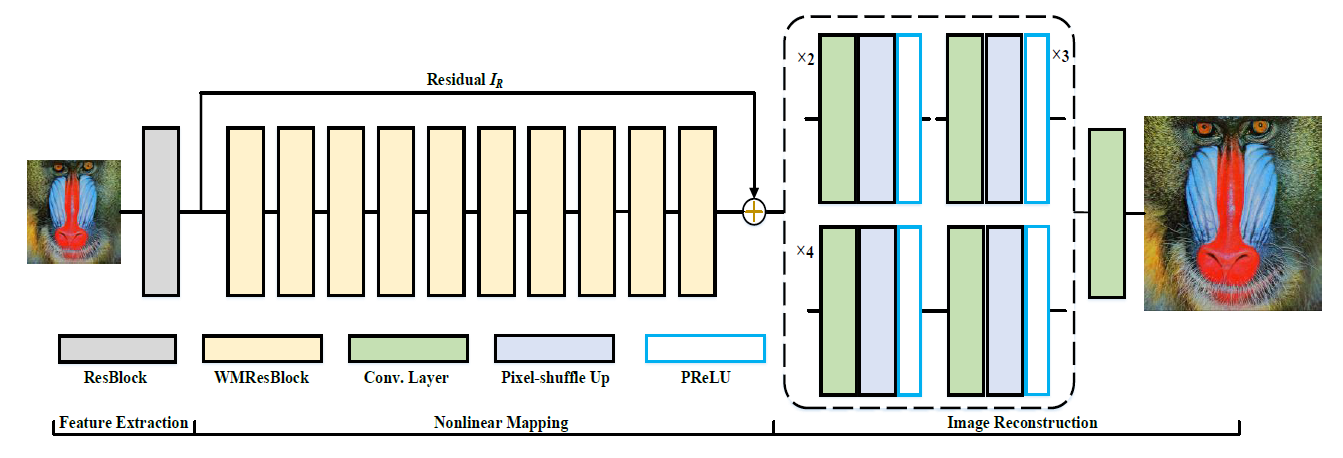}
    \caption{GUET-HMI Team: framework of WMRN model.}
    \label{fig:GUET-HMI-1}
    \vspace{-5mm}
\end{figure}

\begin{figure}[t!]
    \centering
    \includegraphics[width=.9\linewidth]{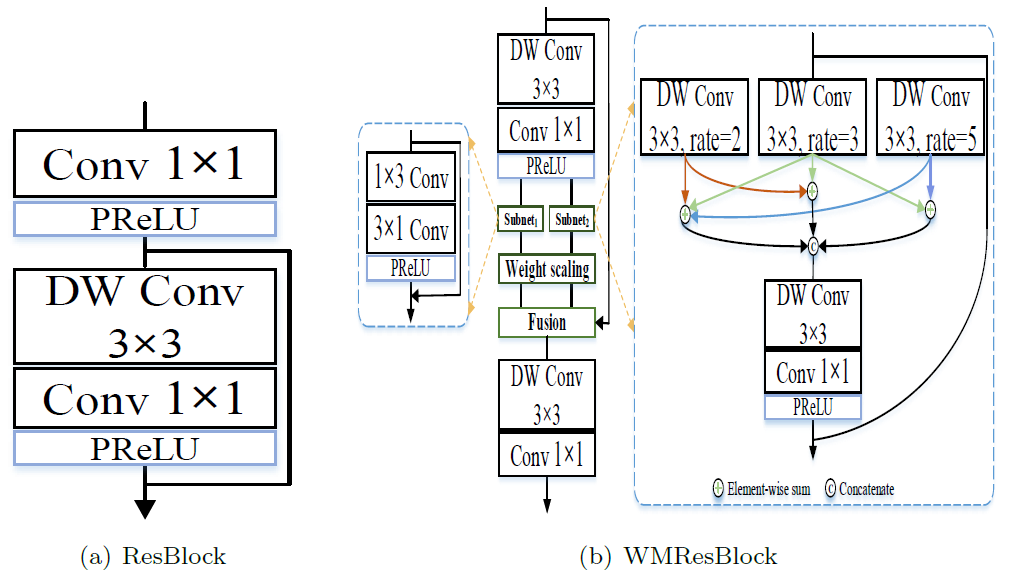}
    \caption{GUET-HMI Team: WMRN building block. (a) ResBlock and (b) WMResBlock.}
    \label{fig:GUET-HMI-2}
    \vspace{-3mm}
\end{figure}

The GUET-HMI team proposed \textbf{weighted multi-scale residual network (WMRN)} (see Figure~\ref{fig:GUET-HMI-1}) for the three tracks. To improve the efficiency, parameter-efficient convolutions (\ie, Depthwise Separable Convolution used in this
work) are extensively utilized to construct the building blocks. WMRN can be divided into three components: feature extraction (FE), nonlinear mapping (NLM) and image reconstruction.
Specifically, WMRN utilizes a residual block (ResBlock) (see Figure~\ref{fig:GUET-HMI-2}(a)) to extract low-level feature information $F_{FE}$. Then, $F_{FE}$ is sent to the nonlinear mapping (NLM) module that contains several WMResBlocks (see Figure~\ref{fig:GUET-HMI-2}(b)) to exploit multi-scale representations of the feature maps.
In order to improve the high frequency details of $I_{SR}$, residual features $I_R$ are additionally added.
Finally, an image recovery sub-net which aims to reconstruct the final high-resolution image $I_{SR}$ is adopted to handle the feature maps of FE and NLM components.
For the training of WMRN, a loss function (\ie, L1 with total variation (TV) regularization) under the assumption that TV penalty could constrain the smoothness of $I_{SR}$ rather than L1 or L2 loss function is adopted.
Denoting $I_{GT}$ as the reference image, the loss function can be written as $L_{total}=\|I_{SR}-I_{GT}\|_1 + \lambda\|\nabla I_{SR}\|_2$,
where $\lambda$ is the balanced weight.

\vspace{-0.2cm}

\paragraph{Acknowledgments. } We thank the AIM 2019 sponsors.

\appendix
\section{Teams and affiliations}
\label{sec:teams}

\subsection*{AIM2019 team}
\noindent\textit{\textbf{Title: }} AIM 2019 Constrained Super-Resolution Challenge\\
\noindent\textit{\textbf{Members: }} \\
Kai Zhang (\href{mailto:kai.zhang@vision.ee.ethz.ch}{kai.zhang@vision.ee.ethz.ch}),\\
Shuhang Gu (\href{mailto:shuhang.gu@vision.ee.ethz.ch}{shuhang.gu@vision.ee.ethz.ch}),\\
Radu Timofte (\href{mailto:radu.timofte@vision.ee.ethz.ch}{radu.timofte@vision.ee.ethz.ch})\\
\noindent\textit{\textbf{Affiliations: }}\\
Computer Vision Lab, ETH Zurich, Switzerland\\

\subsection*{rainbow}
\noindent\textit{\textbf{Title: }}Information Multi-distillation Network (IMDN)\\
\noindent\textit{\textbf{Members: }}\textit{Zheng Hui \\\noindent(\href{mailto:zheng\_hui@aliyun.com}{zheng\_hui@aliyun.com})}, Xiumei Wang, and Xinbo Gao\\
\noindent\textit{\textbf{Affiliation: }}\\
Xidian University
\\

\subsection*{ZJUCSR2019}
\noindent\textit{\textbf{Title: }}NoUCSR: Efficient Neural Network for Super-Resolution without Upsampling Convolution\\
\noindent\textit{\textbf{Members: }}\textit{Dongliang Xiong\\\noindent(\href{mailto:xiongdl@zju.edu.cn}{xiongdl@zju.edu.cn})}\\
\noindent\textit{\textbf{Affiliation: }}\\
College of Electrical Engineering, Zhejiang University
\\

\subsection*{Alpha}
\noindent\textit{\textbf{Title: }}Aggregative Structure in Super Resolution (ASSR)\\
\noindent\textit{\textbf{Members: }}\textit{Shuai Liu$^{1}$
\\\noindent(\href{mailto:935970314@qq.com}{935970314@qq.com)}}, Ruipeng Gang$^{1}$, Nan Nan$^{1}$, Chenghua Li$^{2}$\\
\noindent\textit{\textbf{Affiliation: }}\\
$^{1}$ College of Sciences, North China University of Technology\\
$^{2}$ Institute of Automation, Chinese Academy of Sciences
\\

\subsection*{krahaon\_ai\_cv}
\noindent\textit{\textbf{Title: }}Efficient Super-Resolution Network with Neural Architecture Search\\
\noindent\textit{\textbf{Members: }}\textit{Xueyi Zou \\\noindent(\href{mailto:zouxueyi@huawei.com}{zouxueyi@huawei.com})}, Ning Kang (major contributor), Zhan Wang, Hang Xu\\
\noindent\textit{\textbf{Affiliation: }}\\
Noah's Ark Lab, Huawei
\\

\subsection*{Rookie}
\noindent\textit{\textbf{Title: }}Lightweight Image Super-Resolution with Adaptive Weighted Learning Network\\
\noindent\textit{\textbf{Members: }}\textit{Chaofeng Wang$^{1}$ \\\noindent(\href{mailto:syusuke0516@163.com}{syusuke0516@163.com})}, Zheng Li$^{1}$, Linlin Wang$^{2}$, Jun Shi$^{1}$\\
\noindent\textit{\textbf{Affiliation: }}\\
$^{1}$ Shanghai University\\
$^{2}$ University of California, San Diego
\\

\subsection*{SRSTAR}
\noindent\textit{\textbf{Title: }}DilaResNet for Constrained SR: Fewer Convs, Larger Training Patch Size, and Better Performance\\
\noindent\textit{\textbf{Members: }}\textit{Wenyu Sun \\\noindent(\href{mailto:wy-sun16@mails.tsinghua.edu.cn}{wy-sun16@mails.tsinghua.edu.cn})}\\
\noindent\textit{\textbf{Affiliation: }}\\
Tsinghua University
\\

\subsection*{NPUCS\_103}
\noindent\textit{\textbf{Title: }}Two-stage Recurrent Dense Net\\
\noindent\textit{\textbf{Members: }}\textit{Zhiqiang  Lang \\\noindent(\href{mailto:2015303107lang@mail.nwpu.edu.cn}{2015303107lang@mail.nwpu.edu.cn})}, Jiangtao Nie, Wei Wei, Lei Zhang\\
\noindent\textit{\textbf{Affiliation: }}\\
Northwestern Polytechnical University
\\

\subsection*{PPZ}
\noindent\textit{\textbf{Title: }}Bags of Tricks for the Parameter-Efficient SRResNet\\
\noindent\textit{\textbf{Members: }}\textit{Yazhe Niu
\\\noindent(\href{mailto:niuyazhe@buaa.edu.cn}{niuyazhe@buaa.edu.cn})}, Peijin Zhuo\\
\noindent\textit{\textbf{Affiliation: }}\\
Beihang University; SenseTime (Beijing)
\\

\subsection*{neptuneai}
\noindent\textit{\textbf{Title: }}Lightweight Super Resolution Network with Inverted Residuals Blocks\\
\noindent\textit{\textbf{Members: }}\textit{Xiangzhen Kong
(\href{mailto:neptune.team.ai@gmail.com}{neptune.team.ai@gmail.com})}\\

\subsection*{GUET-HMI}
\noindent\textit{\textbf{Title: }}Towards Real-time Image Super-Resolution via Weighted Multi-scale Residual Network\\
\noindent\textit{\textbf{Members: }}\textit{Long Sun \\\noindent(\href{mailto:lungsuen@163.com}{lungsuen@163.com})}, Wenhao Wang\\
\noindent\textit{\textbf{Affiliation: }}\\
Guilin University of Electronic Technology, China
\\


%
\vspace{-0.4cm}

{\small
\bibliographystyle{ieee_fullname}
\bibliography{egbib}
}

\end{document}